\documentclass[aps,prc,reprint,superscriptaddress,showpacs]{revtex4-1}

\usepackage{graphicx}
\usepackage{dcolumn}
\usepackage{bm}

\begin{document}

\title{On nucleon exchange mechanism in heavy-ion collisions at near-barrier energies}
\author{B. Yilmaz}
\affiliation{Physics Department, Ankara University, Tandogan 06100, Ankara, Turkey}

\author{S. Ayik}
\affiliation{Physics Department, Tennessee Technological University,
Cookeville, Tennessee 38505, USA}

\author{D. Lacroix}
\affiliation{Grand Acc\'el\'erateur National d'Ions Lourds (GANIL), CEA/DSM-CNRS/IN2P3, BP 55027, F-14076 Caen Cedex 5, France}

\author{K. Washiyama}
\affiliation{PNTPM, CP 229, Universit\'e Libre de Bruxelles,
B-1050 Brussels, Belgium }

\date{\today}

\begin{abstract} 
Nucleon drift and diffusion mechanisms in central collisions of asymmetric heavy-ions at near-barrier energies are investigated in the framework of a stochastic mean-field approach. Expressions for diffusion and drift coefficients for nucleon transfer deduced from the stochastic mean-field approach in the semiclassical approximation have similar forms familiar from the phenomenological nucleon exchange model. The variance of fragment mass distribution agrees with the empirical formula $\sigma^2_{AA}(t)= N_{\rm exc}(t)$. The comparison with the time-dependent Hartree-Fock calculations shows that, below barrier energies, the drift coefficient in the semiclassical approximation underestimates the mean number of nucleon transfer obtained in the quantal framework.  Motion of the window in the dinuclear system has a significant effect on the nucleon transfer in asymmetric collisions.   
\end{abstract}

\pacs{25.70.Hi,21.60.Jz,24.60.Ky}

\maketitle

\section{Introduction} 

In heavy-ion collisions with bombarding energies per nucleon in the order of nucleon binding energy, the mean-field approach, in terms of time-dependent Hartree-Fock equations (TDHF), provides a good approximation for describing the average behavior of the  collision dynamics  \cite{Koonin80,Goeke82,Simenel07,Negele82,Davis84}. For example, the mean-field approximation gives a good description of energy dissipation and nucleon drift in deep inelastic heavy-ion collisions (DIC). However, the collective motion is treated in nearly classical, deterministic manner and the fluctuations of collective variables are severely underestimated \cite{Negele82,Davis84}. Therefore, in the mean-field approach, it is not possible to describe energy distributions and fragment mass and charge distributions in DIC. There are other reactions, such as heavy-ion fusion at near-barrier energies, spinodal dynamics leading to nuclear fragmentation \cite{Chomaz}, in which dynamics of density fluctuations play a dominant role. Much work has been done to improve the transport approach for describing dynamics of density fluctuations beyond the mean-field approximation. Basically, there are two different sources of density fluctuations: (i) fluctuations induces by binary collisions 
\cite{Randrup90,Abe96,Lacroix04} and (ii) mean-field fluctuations. Fluctuations and dissipation induced by collisional mechanism are important at the intermediate and high energies. Transport description can be extended in a stochastic approach by including the binary collision term and its fluctuating part in an analogous manner to the Langevin treatment of Brownian motion. In the semiclassical limit, this model is known as the Boltzmann-Langevin approach. On the other hand, as indicated in a recent work \cite{Ayik08}, the mean-field fluctuations originating from the quantal and thermal fluctuations at the initial state become the dominant source of density fluctuations at low energies. The theory that includes mean-field fluctuations is referred to as the Stochastic Mean-Field (SMF) approach.

In recent works, the SMF approach was employed to extract transport coefficients associated with relative momentum and nucleon exchange in low-energy heavy-ion collisions \cite{Ayik09,Washiyama09}. Also, some applications of the SMF approach have been carried out for analyzing the early development of spinodal instabilities \cite{Ayik2008,Ayik2009}. Microscopic transport coefficients extracted from the SMF have similar forms with those familiar from the phenomenological nucleon exchange model \cite{Norenberg74,Randrup78,Chat94}, but they provide a more refined description of one-body dissipation and the associated fluctuation mechanism. In these initial investigations, for simplicity, we calculated transport coefficients for symmetric central collisions. It is worth mentioning that the variances of the fragment mass distributions calculated with the microscopic diffusion coefficient extracted from the SMF approach compares very well with the empirical result  $\sigma^{2}_{AA}(t)=N_{\rm exc}(t)$, where $\sigma^2_{AA}$ and $N_{\rm exc}(t)$ denote the mass variance and the total number of nucleon exchange between projectile and target nuclei up to time $t$, respectively. This observation provides another convincing support for the validity of the SMF approach. In symmetric collisions, the mass and charge numbers of the fragment do not change on the average, i.e., the mean value of nucleon drift coefficient vanishes.  

In this work, we investigate nucleon exchange mechanism in central collisions of asymmetric $^{40}$Ca + $^{90}$Zr system. In Sec. II, we study nucleon drift in the basis of the TDHF and compare the results with those obtained the semiclassical limit of SMF approach. We illustrate the influence of the neck motion on nucleon flux through the window area. In Sec. III, we present a brief description of the SMF approach and transport coefficients for nucleon exchange. In Sec. IV, we investigate the nucleon diffusion and variances of fragment mass distributions, and conclusions are given in Sec. V.  

\section{Nucleon Drift and Window Dynamics} 

As an extension of the previous work, we consider central collisions of asymmetric systems at low energies. For not too heavy systems, collisions above the Coulomb barrier lead to fusion, and at below the barrier energies, colliding nuclei exchange a few nucleons and re-separate. In order to investigate the dynamics of nucleon exchange, we introduce the window between projectile-like and target-like nuclei according to the procedure outlined in 
\cite{Ayik09,Washiyama08}. Figure 1 illustrates density profiles at the reaction plane, $\rho(x,y,z=0,t)$ and the window location in collisions of $^{40}$Ca + $^{90}$Zr system at center-of-mass energy $E_{\rm cm}=97$ MeV at three different times. The Coulomb barrier is $V_B=97.7$ MeV. In this figure and in the rest of this paper, we perform the numerical calculations  with the three-dimensional TDHF code developed by P. Bonche and his collaborators with the SLy4d Skyrme effective force \cite{Kim97}. Since the total nucleon number remains constant, we can take the mass number of the target-like nuclei as the independent variable. 
\begin{figure}[hpt]
\includegraphics[width=8cm]{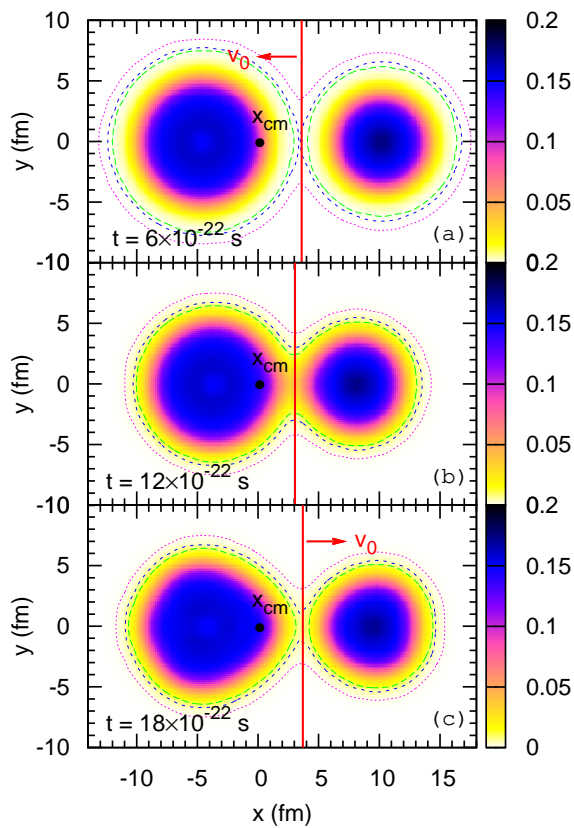}
\caption{Snapshots of the nucleon density profiles on the reaction plane, $\rho(x,y,z=0)$, are indicated by contour plots for the central collision
of $^{40}$Ca + $^{90}$Zr system at $E_{\rm cm}=97$ MeV in units of fm$^{-3}$. The black dot is the center of mass point. The red lines indicate the positions of the window $x_0$ and $v_0=dx_0/dt$ denotes velocity of the window.}
\end{figure}
The mass number of target-like nuclei is determined by integrating the density distribution over the left side of the window,
\begin{equation}
 \label{massproj}
 A_T(t)=\int \int \frac{dx dp_x}{2\pi\hbar}\Theta(x_0-x)f(x,p_x,t),
\end{equation} 
where $x_0(t)$ denotes the position of the neck window. The quantity $\Theta(x_0-x)$ denotes the step function and $f(x,p_x,t)$ is the reduced Wigner function along the collision direction. Definition of the mass number of target-like nucleus is equivalent to
\begin{equation}
 A_T(t)=\int dx \Theta(x_0-x)\rho(x,t),
\end{equation} 
where $\rho(x,t)=\int dp_x/(2\pi\hbar)f(x,p_x,t)$ denotes the reduced density of nucleons.

The Wigner function $f(\boldsymbol{r},\boldsymbol{p},t)$ is defined as a partial Fourier transform of the single-particle density matrix according to
\begin{eqnarray}
f(\boldsymbol{r},\boldsymbol{p},t)&=& \int d^3s\exp\left( - \frac{i}{\hbar}
\boldsymbol{p}\cdot\boldsymbol{s}\right)\rho\left(\boldsymbol{r}+\frac{\boldsymbol{s}}{2},\boldsymbol{r}-\frac{\boldsymbol{s}}{2},t\right),  
\label{eq:wigner}
\end{eqnarray}
where the single-particle density matrix is defined as
\begin{eqnarray}
\label{eq:density}
\rho(\boldsymbol{r},{\boldsymbol{r}}^{\prime},t) = \sum\limits_{i\sigma\tau}
\phi_{i \sigma\tau}^\ast(\boldsymbol{r},t)n_{i}(\sigma\tau)
\phi_{i \sigma\tau}(\boldsymbol{r}^{\prime},t).
\end{eqnarray}
Here, the sum runs over the occupied single-particle wave functions with spin-isospin quantum numbers $\sigma, \tau$ and occupation numbers $n_i$. We obtain the reduced Wigner function $f(x,p_x,t)$ along the collision direction by integrating over the phase space volume on the window area 
between the colliding nuclei,
\begin{eqnarray}
\label{eq:wignerx}
f(x,p_x ,t) = \int dydz\frac{dp_y dp_z}{(2\pi\hbar)^2}
f(\boldsymbol{r},\boldsymbol{p} ,t).
\end{eqnarray}

Using TDHF equations, we can deduce that the rate of change of mass number of target-like fragments as
\begin{eqnarray}
\label{AA}
\frac{d}{dt}A_T(t)=v_A(t),
\end{eqnarray}
where $v_A(t)$ denotes the nucleon drift coefficient determined by the net nucleon flux through the window area,
\begin{eqnarray}
v_A(t) = -\int \frac{dp_x}{2\pi\hbar} \frac{p_x-p_0}{m} f(x_0,p_x ,t).
\label{dAdt}
\end{eqnarray}
Here $p_0=mdx_0/dt$ is the velocity of the neck multiplied by nucleon mass. Even though we employ the reduced Wigner function for convenience, 
the expression of the drift  coefficient is fully quantum mechanical. Since the net nucleon flux is determined by the kinetic term in TDHF, 
Eq. (\ref{dAdt}) does not involve a semiclassical approximation. The reduced Wigner function  $f(x,p_x ,t)$ is exact Wigner transform of the reduced density matrix  $\rho(x+s_{x}/2,x-s_{x}/2 ,t)$. 
The center of mass of the colliding nuclei in Fig. 1 is located at $(x_{cm},y_{cm})=(0,0)$. The red vertical lines indicate the positions of windows
$x_{0}$. As seen from the figure, the position of the window does not remain constant relative to the center of mass. The directions of neck velocities are indicated with arrows during the approach and re-separation phase of nuclei. Figure 2 shows the time dependence of the position of the window relative to the center of mass in collisions of $^{40}$Ca + $^{90}$Zr system at three different center-of-mass energies. As seen from Fig. 2(b), at energies below the Coulomb barrier, during the approach phase, the neck moves towards the center of mass and then moves away from the center of mass. We note that the center-of-mass energy $E_{\rm cm}=110$ MeV leads to fusion. However, as seen in Fig. 2(a), before the system fuses, position of the window undergoes a single oscillation. 
\begin{figure}[hpt]
\includegraphics[width=8cm]{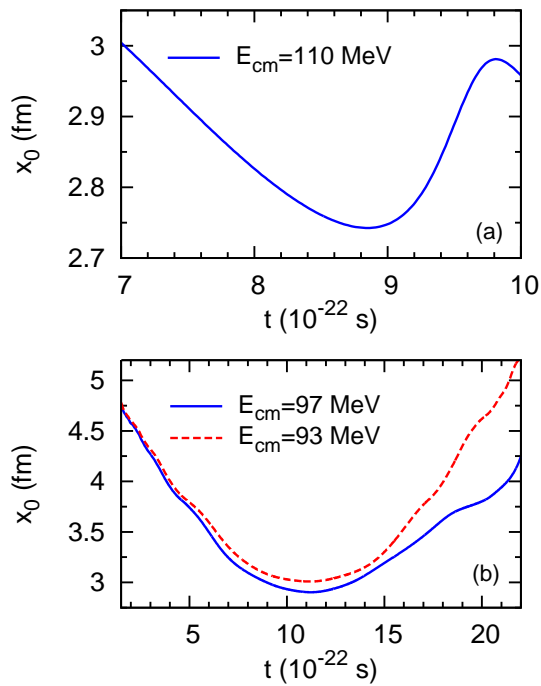}
\caption{(color online) The window position is plotted versus time for the central collisions of $^{40}$Ca + $^{90}$Zr system at three different center-of-mass energies. The center of mass point is located at x~$=0$.}
\end{figure}
\begin{figure}[hpt]
\includegraphics[width=8cm]{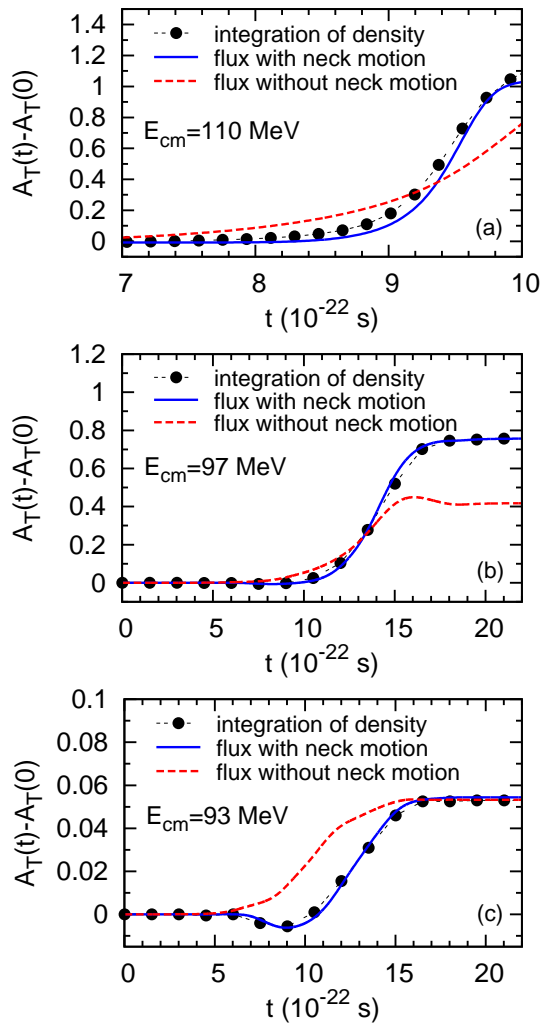}
\caption{(color online) Mean number of nucleon transfer to the target nucleus $^{90}$Zr, $A_T(t)-A_T(0)$, is plotted versus time in central collisions of 
$^{40}$Ca + $^{90}$Zr system at three different center-of-mass energies. The dotted curves are calculated directly from Eq. (\ref{massproj}), the solid lines are obtained by integrating the nucleon flux on the right-hand side of Eq. (\ref{dAdt}) over time, and the dashed lines are  calculated  similarly with solid lines except the motion of the window, second term of the right-hand side of Eq. (\ref{dAdt}), is neglected.}
\end{figure} 

In order to illustrate the effect of window dynamics on nucleon drift, 
in Fig. 3, we show time evolution of the mean mass number of the target-like nuclei, 
\begin{eqnarray}
\label{Ads}
A_T(t)=A_T(0)+ \int_{0}^{t} ds v_A(s),
\end{eqnarray}
as a function of time at three different center-of-mass energies, $E_{\rm cm}=93$ MeV (c), $E_{\rm cm}=97$ MeV (b) and $E_{\rm cm}=110$ MeV (a). In this figure, dotted lines are obtained by integrating the density over the left 
side of the window from Eq. (\ref{massproj}) and the solid lines are obtained by integrating the nucleon flux across the window from Eq. (\ref{dAdt}). Dashed lines are the result of flux calculations from Eq. (\ref{dAdt}), except window motion is neglected. We observe that window motion has an important influence on the net nucleon flux or, in terms of transport language, on the mass drift coefficient, at both below and above barrier energies.  In this connection, we remark that in the phenomenological nucleon exchange model, transport coefficients (mass drift and diffusion) are usually calculated by assuming a fixed window position relative to the center of mass \cite{Randrup78}. As seen in Fig. 3, keeping the window at rest relative to the center of mass can introduce an error in the order of $50\%$ of the mean number of transferred nucleons.

\subsection{Nucleon Exchange in the Stochastic Mean-Field Theory}

In the SMF approach \cite{Ayik08}, the mean-field fluctuations are incorporated into the dynamical evolution by including the initial state fluctuations in a stochastic approximation  in a similar manner to the idea presented in Refs. \cite{Esbensen78,Dasso85,Guidry87,Dasso92,Galetti93,Ayik10}. 
In the approach, in contrast to the deterministic description of the standard mean-field approach, an ensemble of single-particle density matrices are generated by starting an initial density distribution.  A member of the ensemble of single-particle density matrices, denoted by $\lambda$, can be expanded in terms of a complete set of single-particle wave functions as  
\begin{eqnarray}
\label{eq:density2}
\rho^\lambda(\boldsymbol{r},{\boldsymbol{r}}^{\prime},t) = \sum\limits_{ij \sigma\tau}
\phi_{i \sigma\tau}^\ast(\boldsymbol{r},t;\lambda )\rho_{ij}^\lambda(\sigma\tau)
\phi_{j \sigma\tau}(\boldsymbol{r}^{\prime},t;\lambda ),
\end{eqnarray}
where $\sigma$ and $\tau$ indicate the spin-isospin quantum numbers of the single-particle wave functions. 
In the expansion, elements of density matrices are assumed to be random Gaussian numbers with mean values,
\begin{eqnarray}
 \label{mom1}
 \langle\rho_{ij}^\lambda(\sigma\tau)\rangle=\delta_{ij}n_i^{\sigma\tau}
\end{eqnarray}
and variances are determined by
\begin{eqnarray}
\label{mom2}
\langle\delta\rho_{ij}^\lambda(\sigma\tau),
\delta\rho_{j'i'}^\lambda( {\sigma}'{\tau}')\rangle&& \nonumber\\
=\frac{1}{2}\delta_{j{j}'}\delta_{i{i}'}\delta_{\tau {\tau}'}\delta_{\sigma {\sigma}'}&
\left[n_i^{\sigma\tau}(1 - n_j^{\sigma\tau}) + n_j^{\sigma\tau}(1 - n_i^{\sigma\tau})\right]&.
\end{eqnarray} 
Here, $n_i^{\sigma\tau}$ denotes the average single particle occupation factors, which are values 0 and 1 at zero temperature and given by the Fermi-Dirac distribution at finite temperatures. Time evolution of the single-particle wave functions is determined by the self-consistent mean-field 
$h(\rho^\lambda)$ of the corresponding event,
\begin{eqnarray}
\label{eq:spwf}
i\hbar \frac{\partial }{\partial t}\phi_{i \sigma\tau} (\boldsymbol{r},t;\lambda )
= h(\rho^\lambda )\phi_{i \sigma\tau}(\boldsymbol{r},t;\lambda ).
\end{eqnarray}

We introduce the Wigner function $f^\lambda(\boldsymbol{r},\boldsymbol{p},t)$ corresponding to each event $\lambda$ of the ensemble of single-particle density matrices in the same manner as in Eq. (\ref{eq:wigner}). Also, we define the reduced Wigner function $f^{\lambda}(x,p_x,t)$ and the mass number of target-like nuclei $A^{\lambda}_T(t)$ in the same manner as in Eqs. (\ref{eq:wignerx}) and (\ref{massproj}). The time evolution of the mass number of the target-like fragments in the event $\lambda$ obeys the equation
\begin{eqnarray}
\frac{d}{dt}A_T^{\lambda}
= -\int \frac{dp_x}{2\pi\hbar} \frac{p_x-p_0}{m} f^{\lambda}(x_0,p_x ,t),
\label{dAdtt}
\end{eqnarray}
which is similar to Eq. (\ref{AA}) except that the nucleon flux through the window is calculated by the reduced Wigner function in that event.
It is possible to convert this equation into a Langevin equation for stochastic evolution of the mass number of the target-like nuclei. 
Fluctuations of the nucleon flux through the window can arise from fluctuations of the reduced Wigner function originating implicitly from the mass number dependence $\delta A_T^{\lambda}(t)=A_T^{\lambda}(t)-A_T(t)$ and explicitly from the fluctuations of single-particle degrees of freedom
$\delta f^{\lambda}(x_0,p_x,t)$. Then, the reduced Wigner function can be written as 
\begin{eqnarray}
f^{\lambda}(x_0,p_x,t)&=&f(x_0,p_x,t)+\left( \frac{\partial f(x_0,p_x,t)}{\partial A_T}\right) \delta A_T^{\lambda}(t)\nonumber\\
&&+\delta f^{\lambda}(x_0,p_x,t).
\end{eqnarray}
Here, $A_T(t)=\langle A_T^{\lambda}(t)\rangle$ and $f(x_0,p_x,t)=\langle f^{\lambda}(x_0,p_x,t)\rangle$ denote average quantities taken over the generated ensemble.
For small fluctuations, the ensemble average quantities are equivalent to the results obtained by ordinary mean-field approximation. As a result, the
Langevin equation for the nucleon exchange becomes,
\begin{eqnarray}
\label{eq:langevin-mass}
\frac{d}{dt}A_T^\lambda(t) = v_A(t)+\left( \frac{\partial v_A(t)}{\partial A_T} \right)\delta A_T^{\lambda}(t) + \xi_A^\lambda(t),
\end{eqnarray}
where $v_A$ is the drift coefficient for nucleon exchange given by
Eq. (\ref{dAdt}) in the standard mean-field approach. The quantity
$\xi_A^\lambda(t)$ denotes the fluctuating part of the nucleon flux,
\begin{eqnarray}
\xi_A^\lambda(t)=-\int \frac{dp_x}{2\pi\hbar} \frac{p_x-p_0}{m} \delta f^\lambda(x_0,p_x ,t).
\label{vaa}
\end{eqnarray} 
The fluctuating part of the flux is considered as a Markovian Gaussian random force on the nucleon exchange mechanism determined by a correlation function,
\begin{equation}
\langle\xi_A^\lambda(t)\xi_A^\lambda(t')\rangle= 2\delta(t-t')D_{AA}(t),
\end{equation}
where $D_{AA}(t)$ is the diffusion coefficient associated with nucleon exchange. In the SMF approach, within the semiclassical approximation, it is possible to deduce expressions for the diffusion coefficients associated with macroscopic variables \cite{Ayik09,Washiyama09}. The expression of diffusion coefficient for nucleon exchange is given by
\begin{eqnarray}
D_{AA}(t) = \int\frac{dp_x }{2\pi \hbar}
\frac{\vert p_x-p_0 \vert}{m}\frac{1}{2} \Lambda^{+}(x_0,p_x,t),
\label{e1}
\end{eqnarray}
where the quantity $\Lambda^{\pm}(x_0,p_x,t)$ is defined as
\begin{eqnarray}
\Lambda^{\pm}(x_0,p_x,t) &=& \sum_{\sigma\tau} \left\{f_P^{\sigma\tau}(x_0,p_x ,t)
\left[1 - \frac{f_T^{\sigma\tau}(x_0,p_x,t)}{\Omega(x_0,t)}\right]\right.\nonumber \\
&&\;\left.\pm f_T^{\sigma\tau}(x_0,p_x,t)\left[1 - \frac{f_P^{\sigma\tau}(x_0,p_x,t)}{\Omega(x_0,t)}\right]\right\}.
\label{e2}
\end{eqnarray}
\begin{figure}[htbp]
\includegraphics[width=8cm]{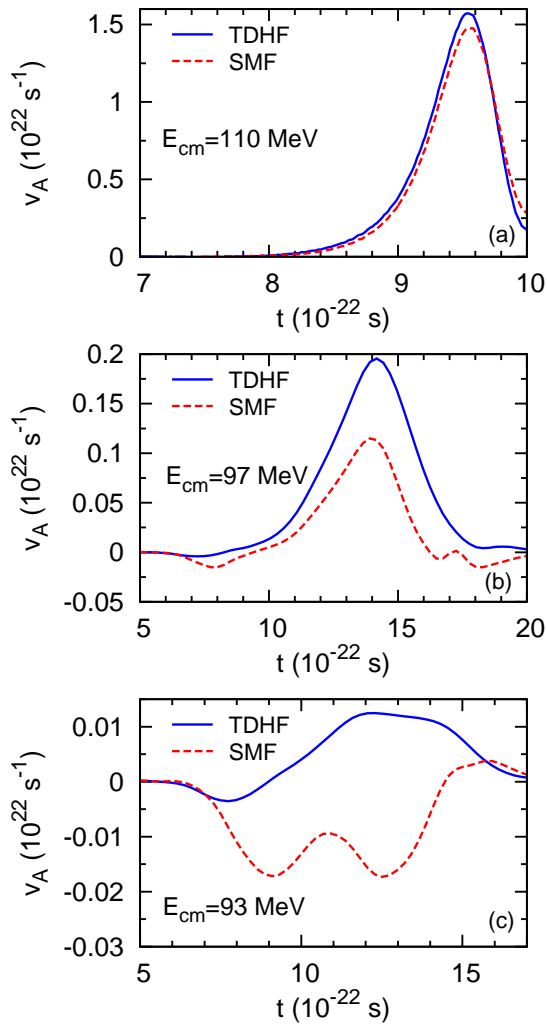}
\caption{(color online) Nucleon drift coefficients are plotted versus time in central collisions of $^{40}$Ca + $^{90}$Zr system at three different center-of-mass energies. The solid lines and dashed lines are obtained from Eq. (\ref{dAdt}) and 
Eq. (\ref{e4}), respectively.}
\end{figure}
\begin{figure}[htbp]
\includegraphics[width=8cm]{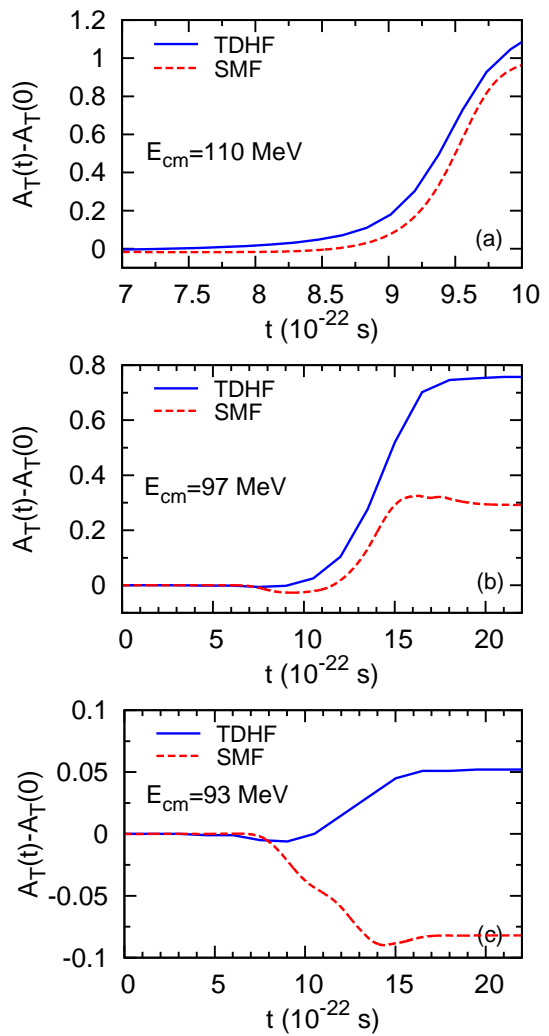}
\caption{(color online) Mean number of nucleon transfer to target nucleus $A_T(t)-A_T(0)$ is plotted versus time in central collisions of  
$^{40}$Ca + $^{90}$Zr system at three different center-of-mass energies. The solid lines are obtained by taking time integral of the TDHF drift from 
Eq. (\ref{dAdt}), and the dashed lines are calculated from the semiclassical expression Eq. (\ref{e4}) of the nucleon drift coefficient.}
\end{figure} 
Here, the mean values of the reduced Wigner distributions $f_{P}^{\sigma\tau}(x_0,p_x,t)$ and $f_{T}^{\sigma\tau}(x_0,p_x,t)$ are defined in terms of single-particle wave functions originating from projectile and target nuclei, respectively. The quantity $\Omega(x_0,t)$ denotes the 
phase-space volume on the window and it is calculated with the method explained in Ref. \cite{Ayik09}. The expression of diffusion coefficient has a similar form with that familiar from the phenomenological nucleon exchange model \cite{Randrup78}. However, diffusion coefficient extracted from the SMF approach is not restricted by adiabatic or diabatic conditions, and provides a more refined description of the nucleon diffusion mechanism.  In the phenomenological nucleon exchange model, diffusion coefficient is determined by the sum of fluxes from projectile to target and from target to projectile, while the nucleon drift coefficient is specified by the net flux through the window. In the spirit of nucleon exchange model, we can infer an approximate expression for the nucleon drift coefficient as 
\begin{eqnarray}
\label{e4}
v_A(t)\approx \int_{-\infty}^{+\infty}\frac{dp_x}{2\pi\hbar}\frac{|p_x-p_0|}{m}\Lambda^{-}(x_0,p_x,t).
\end{eqnarray}
We note that the expressions for the nucleon drift and diffusion coefficients, Eqs. (\ref{e1}) and (\ref{e4}), are valid in the semiclassical approximation. In particular, at below the barrier energies, the reduced Wigner functions $f_{P/T}^{\sigma\tau}(x_0,p_x,t)$ obtained from the TDHF calculations exhibit large oscillations in momentum space. In order to obtain semiclassical description of transport coefficients, we need to smooth out these oscillations in the reduced Wigner functions. Similar to the Gaussian overlap approximation introduced in Ref. \cite{Washiyama09}, we develop a smoothing method of the momentum dependence of the reduced Wigner functions, which is presented in the Appendix.   

\subsection{Comparison between quantal and semiclassical nucleon drift}

Figure 4 shows the drift coefficients as a function of time at center-of-mass energies below the barrier, 
$E_{\rm cm}=93$ MeV (c) and $E_{\rm cm}=97$ MeV (b), and above the barrier energy $E_{\rm cm}=110$ MeV (a).
In the figure, the quantal drift coefficients calculated from Eq. (\ref {dAdt}) and the semiclassical approximation 
of the drift coefficients calculated from Eq. (\ref {e4}) are indicated by solid and dashed lines, respectively. The 
mean-field approach described by TDHF equations provides a quantal treatment of the single particle motion, and 
hence we can take the TDHF drift coefficient as a reference.   As a result, nucleon transfer via barrier penetration is included 
into the quantal drift coefficients. The semiclassical calculations provide a good approximation for the drift coefficients at above 
the barrier energies. However, at below the barrier energies, semiclassical calculations underestimate the quantal drift coefficients progressively more and more. 

Figure 5 shows comparisons of the mean nucleon transfer $A_T(t)-A_T(0)$ calculated from Eq. (\ref{Ads}) in the standard mean-field TDHF and the SMF 
approaches at the same center-of-mass energies below the barrier $E_{\rm cm}=93$ MeV (c) and $E_{\rm cm}=97$ MeV (b), and above the barrier energy 
$E_{\rm cm}=110$ MeV (a). We observe that, 
at above the barrier energies, mean value of the nucleon drift is well described by the semiclassical approximation. However, at below the barrier energies, quantal effects become important, as a result the semiclassical calculations underestimate the quantal value of nucleon drift by more 
than 50\% at $E=97$ MeV.  At $E=93$ MeV, drift in the semiclassical calculation goes in the opposite direction. Figure 6 shows the comparison of asymptotic values of the mean numbers of nucleon transfers in the TDHF and SMF calculations as a function of center-of-mass energy. Below barrier energies, the asymptotic values ($t\rightarrow+\infty$) are the mean numbers of nucleon transfer after re-separation. On the other hand, at the over barrier energies, for the asymptotic values, we take the maximum number of nucleon transfer at the initial stage of the fusion. At below the barrier energies, quantal effects cause the large differences in the mean number of nucleon transfers between the TDHF and the semiclassical calculations.
\begin{figure}[htbp]
\includegraphics[width=8.5cm]{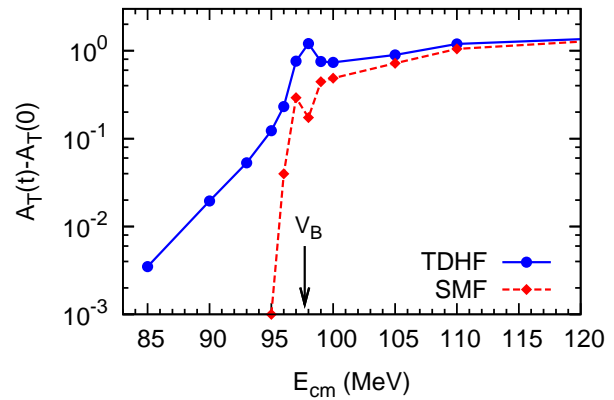}
\caption{(color online) The asymptotic values ($t\rightarrow\infty$) of the mean numbers of nucleon transfer are plotted versus center-of-mass energy 
in collisions of $^{40}$Ca + $^{90}$Zr system. The  Coulomb barrier, $V_{B}$, is indicated by the arrow. At energies above the barrier, where fusion occurs, the maximum numbers of nucleon transfered, for instance the maximum value seen in Fig. 5a, are indicated. The solid lines and  dashed lines are calculated from the standard TDHF Eq. (\ref{massproj}) and the SMF approaches Eq. (\ref{e4}), respectively.}
\end{figure}
\section{Diffusion Coefficient and Variance of Fragment Mass Distribution}

In this section, we calculate the variance, $\sigma^2_{AA}(t)=\langle (A^\lambda_{T})^2\rangle-\langle A^\lambda_{T}\rangle^2$,   
of fragment mass distribution in  central collisions of $^{40}$Ca + $^{90}$Zr system. It follows from the Langevin Eq. (\ref{eq:langevin-mass}), that the variance 
$\sigma^2_{AA}(t)$ is determined by
\begin{equation}
 \frac{d}{dt}\sigma^2_{AA}(t) = 2\alpha(t)\sigma^2_{AA}(t)+ 2D_{AA}(t),
 \label{variance} 
\end{equation} 
where $\alpha(t)=\partial v_A(t)/\partial A_T$. 

\begin{figure}
\includegraphics[width=8cm]{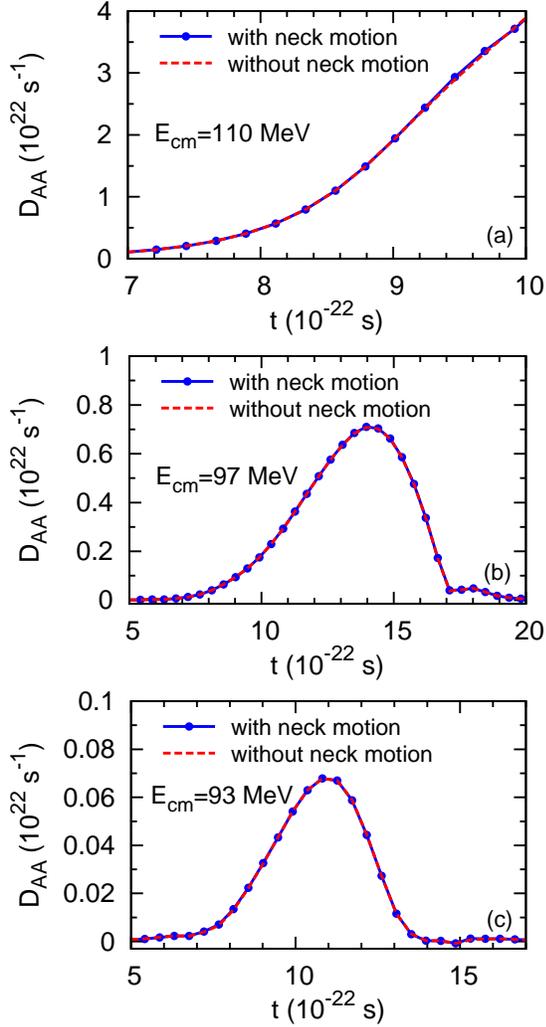}
\caption{(color online) Nucleon diffusion coefficients are plotted versus time in central collisions of $^{40}$Ca + $^{90}$Zr system at three different center-of-mass energies. The solid lines are obtained from Eq. (\ref{e1}), while the dashed lines are calculated by neglecting motion of the window.}
\end{figure}

Solid lines in Fig. 7 illustrate the diffusion coefficients as a function of time at center-of-mass energies below the barrier, 
$E_{\rm cm}=93$ MeV (c) and $E_{\rm cm}=97$ MeV (b), and above the barrier energy $E_{\rm cm}=110$ MeV (a). The solid lines are obtained by Eq. (\ref{e1}), while the dashed lines are calculated by neglecting motion of the window. We observe from this figure that, contrary to behavior of drift coefficients, the motion of the window does not affect diffusion coefficients appreciably. Time dependence of drift coefficients $v_{A}(t)$  in central collisions of asymmetric 
$^{40}$Ca + $^{90}$Zr system at three different center-of-mass energies is illustrated in Fig. 4. In particular, at below the barrier energies, time dependence of drift coefficients in both TDHF and SMF approaches arises mainly from time dependence of the window area and time dependence of nucleon density on the window area. Furthermore, because of very small value of the mean nucleon transfer, 
mass asymmetry degree of freedom is far from equilibrium. Therefore, the reduced mass dependence of the drift coefficients is expected to be very small.
In the following, we neglect the contribution from drift term and solve the variance equation (\ref{variance}) to find, 
\begin{equation}
\sigma^2_{AA}(t) = 2 \int_0^t D_{AA}(s)ds.
\label{eq:sigma}
\end{equation} 
Figure 8 shows variances of fragment mass distribution at the same center-of-mass energies below the barrier $E_{\rm cm}=93$ MeV (c) and $E_{\rm cm}=97$ MeV (b), and above the barrier energy $E_{\rm cm}=110$ MeV (a), as a function of time. Solid lines in the figure are the result of integration over the diffusion coefficients, while solid dots indicate the total number of exchange nucleon $N_{\rm exc}(t)$ until time $t$. 
The empirical relation $\sigma^2_{AA}(t)= N_{\rm exc}(t)$ follows from the nucleon exchange model and has been widely used to analyze the experimental data of deep-inelastic collisions \cite{Freiesleben,Adamian}. Figure 8 illustrates, as in the previous investigations for symmetric heavy-ion collisions \cite{Washiyama09}, that there is a good agreement between the empirical formulae and  the calculations based on the SMF approach. Even at energies below the Coulomb barrier,  the semiclassical calculations based on the SMF approach compares well with 
the empirical description $N_{\rm exc}$. 
\begin{figure}[hpt]
\includegraphics[width=8cm]{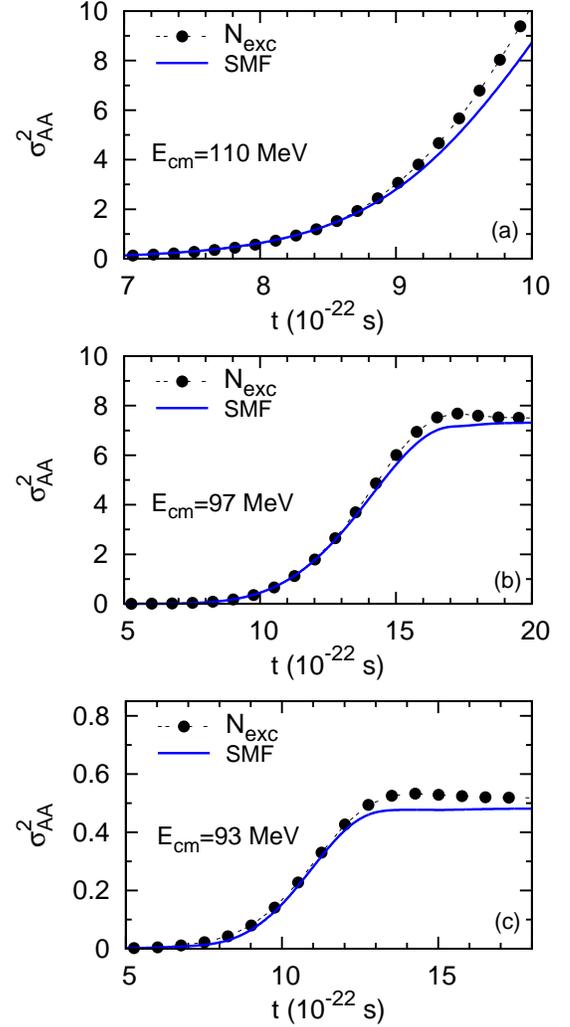}
\caption{(color online) Variances of fragment mass distributions are plotted versus time in collisions of  $^{40}$Ca + $^{90}$Zr system at three different center-of-mass energies. The dotted lines denote total number of exchanged nucleons until a given time $t$, while the solid lines are calculated 
from Eq. (\ref{eq:sigma}).}
\end{figure} 

\section{Conclusion}

We investigate nucleon drift and diffusion mechanisms in central collisions of asymmetric $^{40}$Ca + $^{90}$Zr system in frameworks of the standard mean-field and the SMF approaches.  In the SMF approach, we do not carry out stochastic mean-field calculations, but we extract diffusion coefficient associated with nucleon exchange in a semiclassical approximation, which has a form similar to that familiar from the phenomenological nucleon exchange model. Using this similarity, we also infer an expression for the nucleon drift coefficient in the semiclassical approximation.  In the standard mean-field approach provided by the TDHF equations, we can only calculate the nucleon drift including quantum mechanical effects in the single-particle level, but we can not describe nucleon diffusion in this framework. At energies below the Coulomb barrier, colliding nuclei exchange a few particles and re-separate. At above the barrier energies, since collisions lead to fusion, we limit our calculations of nucleon exchange during the early stages of the collisions. Our calculations show that not only the size of window between nuclei but also motion of window relative to the center of mass play an important role in nucleon drift mechanism. In the phenomenological nucleon exchange model, the motion of the window relative to the center of mass is often neglected. We observe that at below the barrier energies, the semiclassical drift coefficient underestimates the nucleon drift deduced from the standard mean-field description with TDHF equations. This is mainly due to the fact that barrier penetration of the single-particles is not included in the semiclassical description of nucleon drift. However, in collisions with energies above the Coulomb barrier, the semiclassical expression for the nucleon drift provide a good approximation for the quantal drift deduced from the TDHF framework. Employing the diffusion coefficient extracted from the SMF approach, we calculate the variances  of fragment mass distribution at energies below and above the barrier. In such calculations for central collisions, we cannot compare the results with experiments, but can compare them with the empirical relation 
$\sigma^2_{AA}(t)= N_{\rm exc}(t)$. As in the symmetric collisions we investigated earlier \cite{Ayik09,Washiyama09}, our calculations with semiclassical diffusion coefficients  for asymmetric system compare well with the empirical relation at energies both below and above the Coulomb barrier. This empirical result has been used to analyze data in DIC above the barrier energies \cite{Freiesleben,Adamian}. However, it is not clear whether such empirical relation $\sigma^2_{AA}(t)= N_{\rm exc}(t)$ is also valid below the Coulomb barrier, where quantal effects may become important in nucleon exchange mechanism, as we observe in nucleon drift coefficients. Therefore, in the present calculations, we cannot draw any conclusions about the possible quantal effects on nucleon diffusion coefficients and the resulting variances of fragment mass distributions at energies below the Coulomb barrier.  

\begin{acknowledgments} 
We thank P. Bonche for providing the 3D-TDHF code. S.A., B.Y., and K.W. gratefully acknowledge GANIL for the support and warm hospitality extended to them during their visits. This work is supported in part by the US DOE Grant No. DE-FG05-89ER40530.
\end{acknowledgments}

\appendix*
\section{}

We consider the reduced Wigner function
\begin{eqnarray}
 f_{T}^{\sigma\tau}(x_0,p_x ,t) &=& \int dydz
\int ds_x\exp\left( - \frac{i}{\hbar}p_x s_x\right)\nonumber \\
&&\times \sum_{i\in T}
\phi_{i \sigma\tau}^\ast \left(x + \frac{s_x}{2},y,z,t\right)\nonumber\\
&&\quad\quad\times\phi_{i \sigma\tau}\left(x - \frac{s_x}{2},y,z,t\right),
\label{eq:wignerPT}
\end{eqnarray}
which is associated with the single-particle wave functions of the target nucleus. It can be rewritten as
\begin{eqnarray}
 f(x_0,p)=\int ds \exp(-ips/\hbar)F(x_0,s),
\end{eqnarray}   
where
\begin{eqnarray}
 F(x_0,s)&=&\int dydz\sum_{i}\phi^*_i(x_0+\frac{s}{2},y,z)\nonumber\\
&&\qquad\qquad\;\times\phi_i(x_0-\frac{s}{2},y,z).
\label{a3}
\end{eqnarray}
In these expressions for simplicity, we introduce the notation $f(x_0,p)\equiv f_{T}^{\sigma\tau}(x_0,p_x,t)$ and 
$\phi_i(x_0,y,z)\equiv\phi_{i \sigma\tau}\left(x_0,y,z,t\right)$ with single-particle 
wave functions originating from the target nucleus. At energies below the Coulomb barrier, 
the reduced Wigner function $f(x_0,p)$ exhibits rapid oscillations on the window and can take negative values 
due to quantal effects. In order to 
smooth out these oscillations, we need to approximate the quantity $F(x_0,s)$ in terms of a smooth function as a 
function of $s$, which should be consistent with the following two constraints,
\begin{itemize}
 \item  $F^*(x_0,s)=F(x_0,-s)$, hence the real part must be an even function and the imaginary part must be odd function.
 \item  $\lim_{s\rightarrow\infty}F(x_0,s)= 0$.   
\end{itemize}
A possible choice consistent with these requirements is to approximate $F(x_0,s)$ in terms of a Gaussian function. In order to
determine the centroid and the width of the Gaussian representation, in Eq. (\ref{a3}) we expand the single-particle wave functions 
around $s=0$ up to the second order to find,
\begin{equation}
\label{eq:Fs}
 F(x_0,s)\approx \overline{\rho}(x_0)\left(1-i\beta s-\frac{\alpha}{4} s^2\right).
\end{equation} 
Here, the quantities $\alpha$ and $\beta$ are given by
\begin{eqnarray}
 \beta&=&\frac{\overline{j_x}(x_0)}{\overline{\rho}(x_0)},\\
 \alpha&=&\frac{1}{\overline{\rho}(x_0)}\left[\overline{p^2_x}(x_0)+\overline{\tau_x}(x_0)\right].
\end{eqnarray} 
In these expressions, the reduced mass density, the current density, the squared-momentum density, 
and the kinetic energy density are defined as 
\begin{eqnarray}
\overline{\rho}(x)&=&\int dydz \sum_i\phi_i^*(x,y,z)\phi_i(x,y,z),\\
\overline{j_x}(x)&=&\int dydz \frac{1}{2i}\sum_i(\phi^*_i\bigtriangledown_x\phi_i-\phi_i\bigtriangledown_x\phi^*_i),\\
\overline{p^2_x}(x)&=&-\frac{1}{2}\int dydz\sum_i(\phi^*_i\bigtriangledown^2_x\phi_i+\phi_i\bigtriangledown^2_x\phi^*_i),\label{p2x}\\
\overline{\tau_x}(x)&=&\int dydz\sum_i|\bigtriangledown_x\phi_i|^2.
\end{eqnarray}
In mean-field calculations, the expectation value of kinetic energy is computed by using either the reduced kinetic energy density 
$\overline{\tau_x}(x)$ or the reduced squared-momentum density $\overline{p^2_x}(x)$, since these quantities satisfy a sum rule relation, 
\begin{equation}
 \int_{-\infty}^{\infty} \overline{p^2_x}(x)dx = \int_{-\infty}^{\infty} \overline{\tau_x}(x)dx.
\end{equation} 
We note that the reduced kinetic energy density, $\overline{\tau_x}(x)$, is positive for all $x$ values. However, the reduced 
squared-momentum density, $\overline{p^2_x}(x)$, can take small negative values in the vicinity of the window below barrier energies. 
As an example, Fig. 9 shows the quantities $\overline{\tau_x}(x)$ and $\overline{p^2_x}(x)$ in central collisions of
$^{90}$Zr and $^{40}$Ca system at a center-of-mass energy of  $E_{\rm cm}=97 $ MeV and at time  $t=10.5\times10^{-22}$ s. 
We observe that the quantity $\overline{p^2_x}(x)$ for target as well as projectile can take small negative value in the
vicinity of window position $x_0$. Consequently, the parameter $\alpha$ can take negative values. 
\begin{figure}[hpt]
\includegraphics[width=8cm]{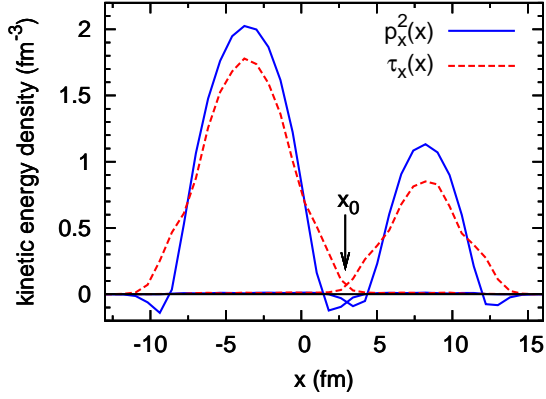}
\caption{Quantities $\overline{p^2_x}(x)$ (solid line) and $\overline{\tau_x}(x)$ (dashed line) are plotted versus position $x$ 
in collisions of  $^{40}$Ca + $^{90}$Zr system 
at $E_{\rm cm}=97 $ MeV and at time  $t=10.5\times10^{-22}$ s for target and projectile. 
The neck position $x_0$ is indicated by an arrow on the figure.}
\end{figure}

In the Gaussian approximation, in order to produce the correct second order Taylor expansion,  we introduce a new parameter $\gamma = \alpha-2\beta^2$.
When $\alpha>0$ and $\gamma>0$, we can approximate the quantity $F(x_0,s)$ in terms of a single Gaussian as
\begin{equation}
  F(x_0,s)\approx\overline{\rho}(x_0)\exp\left(-i\beta s-\frac{\gamma}{4} s^2\right),
\end{equation} 
By taking the Fourier transform, we obtain a Gaussian form for the reduced Wigner function as well,
\begin{equation}
\label{asd}
 f(x_0,p)=2\overline{\rho}(x_0)\sqrt{\frac{\pi}{\gamma}}\exp\left[-\frac{(p-\hbar\beta)^2}{\hbar^2\gamma}\right].
\end{equation} 
When the quantity $\alpha<0$, we can write the expansion Eq. (\ref{eq:Fs}) as
\begin{equation}
\label{eq:Fsn}
 F(x_0,s)\approx\overline{\rho}(x_0)\left(1-i\beta s+\frac{|\alpha|}{4}s^2\right).
\end{equation}    
There are many possibilities to approximate $F(x_0,s)$ in terms of Gaussian functions that have the same second order Taylor expansion with Eq. (\ref{eq:Fsn}) and that are consistent with the two constraints indicated above. We consider a double Gaussian approximation according to      
\begin{eqnarray}
F(x_0,s)&=&\overline{\rho}(x_0)\exp\left(-i\beta s\right)\nonumber\\
&&\times\left[2\exp(-\frac{|\gamma|}{8} s^2)-\exp(-\frac{|\gamma|}{2} s^2)\right],
\end{eqnarray}
which leads to the Wigner function given by
\begin{eqnarray}
 f(x_0,p)&=&\overline{\rho}(x_0)\sqrt{\frac{2\pi}{|\gamma|}}\left\{4\exp\left[-\frac{2(p-\hbar\beta)^2}{\hbar^2|\gamma|}\right]\right.\nonumber\\
 &&\qquad\qquad\left.-\exp\left[-\frac{(p-\hbar\beta)^2}{2\hbar^2|\gamma|}\right]\right\}.
\label{asd2}
\end{eqnarray}
Figure 10 shows an illustration of the reduced Wigner functions given by Eqs. (\ref{asd}) and (\ref{asd2}) by employing the same $\beta$ and the same magnitude of $\gamma$. We observe that for $\gamma<0$, the reduced Wigner function exhibits a negative tail, which is closely related to the quantal behavior of the Wigner function. In the calculations presented in this work, relatively small negative tails in the reduced Wigner functions do not make a significant effect on the semiclassical expressions of the nucleon diffusion and drift coefficients given by Eqs. (\ref{e1}) and (\ref{e4}). 
\begin{figure}[hpt]
\includegraphics[width=8cm]{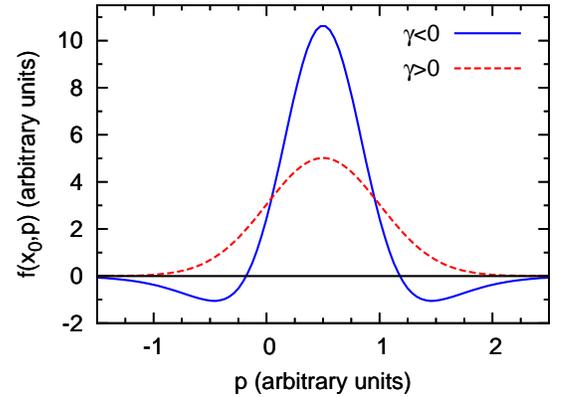}
\caption{Reduced Wigner functions, Eq. (\ref{asd}) and Eq. (\ref{asd2}), are plotted in arbitrary units with parameters 
$\overline{\rho}(x_0)=1$, $\beta=0.5$, and $\gamma=\pm 0.5$. }
\end{figure}
Substituting Eq. (\ref{asd}) and/or (\ref{asd2})  for the reduced Wigner functions originating from target and projectile into Eq. (\ref{e1}), we obtain the nucleon diffusion coefficient as
\begin{eqnarray}
 D_{AA}&=&\Gamma(\overline{\rho}_P,\beta_P,\gamma_P)+\Gamma(\overline{\rho}_T,\beta_T,\gamma_T)\nonumber\\
&&+\Pi(\gamma_P,\gamma_T),
\end{eqnarray}
for $\gamma_P>0$ and $\gamma_T>0$,
\begin{eqnarray}
D_{AA}&=&2\Gamma(\overline{\rho}_P,\beta_P,|\gamma_P|/2)-\Gamma(\overline{\rho}_P,\beta_P,2|\gamma_P|)\nonumber\\ 
&&+\Gamma(\overline{\rho}_T,\beta_T,\gamma_T)+2\Pi(|\gamma_P|/2,\gamma_T)\nonumber\\
&&-\Pi(2|\gamma_P|,\gamma_T),
\end{eqnarray}
for $\gamma_P<0$ and $\gamma_T>0$, and
\begin{eqnarray}
&&D_{AA}=2\Gamma(\overline{\rho}_P,\beta_P,|\gamma_P|/2)-\Gamma(\overline{\rho}_P,\beta_P,2|\gamma_P|)\nonumber\\ 
&&+2\Gamma(\overline{\rho}_T,\beta_T,|\gamma_T|/2)-\Gamma(\overline{\rho}_T,\beta_T,2|\gamma_T|)\nonumber\\ 
&&+4\Pi(|\gamma_P|/2,|\gamma_T|/2)+\Pi(2|\gamma_P|,2|\gamma_T|)\nonumber\\
&&-2\Pi(|\gamma_P|/2,2|\gamma_T|)-2\Pi(2|\gamma_P|,|\gamma_T|/2),
\end{eqnarray}
for $\gamma_P<0$ and $\gamma_T<0$. The functions $\Gamma$ and $\Pi$ are defined as
\begin{equation}
 \Gamma(\overline{\rho},\beta,\gamma)=\frac{\hbar}{2m}\overline{\rho}
\left[\sqrt{\frac{\gamma}{\pi}}\exp\left( -\frac{\beta^2}{\gamma} \right)+\beta
\,\mbox{erf}\left( \frac{\beta}{\sqrt{\gamma}} \right)\right],
\end{equation} 
\begin{eqnarray}
 &&\Pi(\gamma_P,\gamma_T)=-\frac{2\hbar}{m}\frac{\overline{\rho}_P\overline{\rho}_T}{\Omega}
\frac{\sqrt{\gamma_P\gamma_T}}{\gamma_P+\gamma_T}\exp\left(-\frac{\beta^2_P}{\gamma_P}-\frac{\beta^2_T}{\gamma_T}\right)\nonumber\\
&&\times\left[1+\frac{\sqrt{\pi}(\gamma_P\beta_T+\gamma_T\beta_P)}{\sqrt{\gamma_P\gamma_T(\gamma_P+\gamma_T)}}\exp\left(\frac{(\gamma_P\beta_T+\gamma_T\beta_P)^2}{\gamma_P\gamma_T(\gamma_P+\gamma_T)}\right)\right.\nonumber\\
&&\qquad\left.\times\,\mbox{erf}\left(\frac{\gamma_P\beta_T+\gamma_T\beta_P}{\sqrt{\gamma_P\gamma_T(\gamma_P+\gamma_T)}}\right)\right], 
\end{eqnarray} 
where the error function is given by
\begin{equation}
 \mbox{erf}(x)=\frac{2}{\sqrt{\pi}}\int_0^x e^{-y^2}dy.
\end{equation} 
Employing the formula given in Ref.\cite{Ayik09} for the phase space volume $\Omega(x_0,t)$ on the window, we obtain for following expressions for the nucleon drift coefficient, Eq. (\ref{e4}), 
\begin{eqnarray}
 v_A=\Gamma(\overline{\rho}_P,\beta_P,\gamma_P)-\Gamma(\overline{\rho}_T,\beta_T,\gamma_T)
\end{eqnarray}
for $\gamma_P>0$ and $\gamma_T>0$,
\begin{eqnarray}
 v_A&=&2\Gamma(\overline{\rho}_P,\beta_P,|\gamma_P|/2)-\Gamma(\overline{\rho}_P,\beta_P,2|\gamma_P|)\nonumber\\ 
&&-\Gamma(\overline{\rho}_T,\beta_T,\gamma_T)
\end{eqnarray}
for $\gamma_P<0$ and $\gamma_T>0$, and 
\begin{eqnarray}
 v_A&=&2\Gamma(\overline{\rho}_P,\beta_P,|\gamma_P|/2)-\Gamma(\overline{\rho}_P,\beta_P,2|\gamma_P|)\nonumber\\ 
&&-2\Gamma(\overline{\rho}_T,\beta_T,|\gamma_T|/2)+\Gamma(\overline{\rho}_T,\beta_T,2|\gamma_T|) 
\end{eqnarray}
for $\gamma_P<0$ and $\gamma_T<0$.


\begin{thebibliography}{99}
\bibitem{Koonin80} S. E. Koonin, Prog. Part. Nucl. Phys. {\bf 4}, 283 (1980).
\bibitem{Goeke82} K. Goeke and P.-G. Reinhard, {\it Time-Dependent Hartree-Fock and Beyond} (Bad
Honnef, Germany, 1982).
\bibitem{Simenel07} C. Simenel, D. Lacroix, and B. Avez, {\it Quantum Many-Body Dynamics: Applications to Nuclear Reactions}, (VDM Verlag, Germany, 2010).
\bibitem{Negele82} J. W. Negele, Rev. Mod. Phys. {\bf 54}, 913 (1982).
\bibitem{Davis84} K. T. R. Davies, K. R. S. Devi, S. E. Koonin, and M. R. Strayer, in {\it Treatise in
Heavy-Ion Science}, ed. D. A. Bromley (Plenum, New York, 1984), Vol. 4.
\bibitem{Chomaz} Ph. Chomaz, M. Colonna and J. Randrup, Phys. Rep. {\bf 389}, 263 (2004).
\bibitem{Randrup90} J. Randrup and B. Remaud, Nucl. Phys. {\bf A514}, 339 (1990).
\bibitem{Abe96} Y. Abe, S. Ayik, P.-G. Reinhard, and E. Suraud, Phys. Rep. {\bf 275}, 49 (1996).
\bibitem{Lacroix04} D. Lacroix, S. Ayik, and Ph. Chomaz, Prog. Part. Nucl. Phys. {\bf 52}, 497 (2004).
\bibitem{Ayik08} S. Ayik, Phys. Lett. {\bf B658}, 174 (2008).
\bibitem{Ayik09} S. Ayik, K. Washiyama, and D. Lacroix, Phys. Rev. C {\bf 79}, 054606 (2009).
\bibitem{Washiyama09} K. Washiyama, S. Ayik, and D. Lacroix, Phys. Rev. C {\bf 80}, 031602(R) (2009).
\bibitem{Ayik2008} S. Ayik, N. Er, O.Yilmaz, and A. Gokalp, Nucl. Phys. A {\bf 812}, 44 (2008).
\bibitem{Ayik2009} S. Ayik, O. Yilmaz, N. Er, A. Gokalp, and P. Ring, Phys. Rev. C {\bf 80}, 034613 (2009). 
\bibitem{Norenberg74} W. N\"orenberg, Phys. Lett. {\bf B53}, 289 (1974).
\bibitem{Randrup78} J. Randrup, Nucl. Phys. {\bf A307}, 319 (1978); Nucl. Phys. {\bf A327}, 490 (1979); Nucl. Phys. {\bf A383}, 468 (1982).
\bibitem{Chat94} S. Chattopadhyay and D. Pal, J. Phys. G {\bf 20}, 357 (1994).
\bibitem{Washiyama08} K. Washiyama and D. Lacroix, Phys. Rev. C {\bf 78}, 024610 (2008).
\bibitem{Kim97} K.-H. Kim, T. Otsuka, and P. Bonche, J. Phys. G {\bf 23}, 1267 (1997).
\bibitem{Esbensen78} H. Esbensen, A. Winther, R. A. Broglia, and C. H. Dasso, Phys. Rev. Lett. {\bf 41}, 296 (1978).
\bibitem{Dasso85} C. H. Dasso, Proc. of the la Rabida Int. Summer School on Nuclear Physics, p. 398, eds. M. Lozano and G. Madurga (World Scientific, Singapore, 1985). 
\bibitem{Guidry87} M. W. Guidry, R. Donangelo, J. O. Rasmussen, and M. S. Hussein, Phys. Rev. C {\bf 36}, 609 (1987). 
\bibitem{Dasso92} C. H. Dasso and R. Donangelo, Phys. Lett. {\bf B276}, 1 (1992). 
\bibitem{Galetti93} D. Galetti, A. Eiras, F. F. deSouzaCruz, J. R. Marinelli, and M. M. W. deMoraes, Phys. Rev. C {\bf 48}, 3131 (1993). 
\bibitem{Ayik10} S. Ayik, B. Yilmaz, and D. Lacroix, Phys. Rev. C {\bf 81}, 034605 (2010).
\bibitem{Freiesleben} H. Freiesleben and J. V. Kratz, Phys. Rep. {\bf 106}, 1 (1984).
\bibitem{Adamian} G. G. Adamian, A. K. Nasirov, N. V. Antonenko, and R. V. Jolos, Phys. Part. Nucl. {\bf 25}, 583 (1994). 
\end{thebibliography}
\end{document}